\documentclass[preprint,12pt]{elsarticle}
\usepackage{graphicx,t1enc}
\usepackage{epsfig}
\usepackage{amssymb,amsmath}
\usepackage[utf8]{inputenc}

\journal{Physica A}

\begin{document}
\begin{frontmatter}
 
\title{Effects of individual attitudes and motion decisions in room evacuation models}

\author[ad1,ad2,ad3]{V. Dossetti}
\author[ad4,ad5]{S. Bouzat}
\author[ad4,ad5]{M. N. Kuperman}
\address[ad1]{CIDS-Instituto de Ciencias, Benem\'erita Universidad Aut\'onoma de 
Puebla, Av. San Claudio esq. 14 Sur, Edif. 103D, Puebla, Pue. 72570, Mexico}
\address[ad2]{Instituto de F\'isica, Benem\'erita Universidad Aut\'onoma de 
Puebla, Apdo.\ Postal J-48, Puebla 72570, Mexico}
\address[ad3]{Consortium of the Americas for Interdisciplinary Science, 
University of New Mexico, Albuquerque, NM 87131, USA}
\address[ad4]{Consejo Nacional de Investigaciones Cient\'{\i}ficas y T\'ecnicas}
\address[ad5]{Centro At\'omico Bariloche (CNEA), (8400) Bariloche, R\'{\i}o
Negro, Argentina}


\begin{abstract}
In this work we present a model for the evacuation of pedestrians from an 
enclosure considering a continuous space substrate and discrete time. 
We analyze the influence of behavioral features that affect the  use of the empty space, 
that can be linked to the attitudes or characters of the pedestrians. 
We study how the interaction of different behavioral profiles affects the needed time to  
evacuate completely a room and the occurrence of clogging. We find that
neither fully egotistic nor fully cooperative attitudes are optimal from the point of view of the 
crowd. In contrast, intermediate behaviors provide lower evacuation times. This lead us to identify 
some phenomena closely analogous to the
{\em faster-is-slower} effect. The proposed model enables for the introduction of Game 
Theory elements to solve conflicts between pedestrians which try to occupy the same space. 
Moreover, it allows for distinguishing between the role of the attitudes in the search for empty 
space and the attitudes in the conflicts. In the present version we only focus on the first of 
these instances. 
\end{abstract}

\end{frontmatter}

\section{Introduction}

During the last decade there has been a multiplication of mathematical models  
for pedestrian dynamics and enclosure evacuation aiming at unveiling the 
most relevant aspects affecting such processes. Despite the fact that all these 
models represent similar processes (or even the same), they implement different methodologies 
with their corresponding  strengths and limitations.

The consideration of  behavioral aspects related to the interactions between individuals 
has produced some notable contributions in the area \cite{hel1,hel2}. In particular, using 
a mechanistic approach, the so called \emph{social-force} model \cite{hel2, par05, lak05, par07} 
captures the fact that the motion of an agent is governed by the desire of 
reaching a certain destination and by the influence it suffers from the 
environment, which includes the other agents. The model links the collective 
motion of the individuals to self-driven many particle systems (see \cite{dos15} and 
references therein). In order to simplify the calculations, advances have also been maid with 
cellular automaton models \cite{kir03, wei06}. 

Recently, some attention was drawn upon the fact that a full description of the behavioral 
aspects involved in an evacuation process needs to consider the disposition of the individuals 
as internal states which may influence the responses of the pedestrians and, ultimately, 
their motion (see, for example, \cite{nic16}). The idea is that not only the environment 
but also the internal motivations govern their reactions \cite{note1}. Thus, some of the 
interactions 
between two or more individuals in a moving crowd can be thought as conflicts which the pedestrians can 
tackle using different strategies, according to their characters or attitudes. These considerations
have naturally led to the inclusion of Game Theory elements in the models for pedestrian dynamics.
The first works considering Game Theory within evacuation models \cite{bro,col} analyzed the 
emergence of pushing behavior in evacuation situations. Subsequently, several authors analyzed various
aspects of the complex subject of linking individual nature to crowd dynamics 
\cite{heli1,lo,heli2,hao,shi,bou,wang} using Game Theory. All these works considered two available 
strategies for each pedestrian. Namely, a cooperative one and a defecting one. The cooperative 
strategy is normally associated to kind and patient behavior consisting in not looking for taking 
advantage from others, and, specially, not rushing to an empty space when a favorable move is 
available  \cite{heli1,bou}. 
Conversely, the defective strategy can represent an impatient pedestrian, who rushes to conquer 
gaps, disrespectful to others. 

In a slightly different context, it was suggested recently \cite{mous} that, despite the success of 
the models based on the Social forces paradigm, more simple assumptions could lead to highly 
satisfactory predictions of pedestrian behavior.
In \cite{mous}, the authors consider that the reaction of an individual is mostly governed by the 
visually perceived environment, and that a pedestrian, instead of being repelled by the neighbors, 
actively looks for empty space or free paths through the crowd. Hence, from different points of 
view, the attitude of the individuals toward the empty space 
(i.e. the way in which pedestrians use and let use the available space) arise as a 
character-dependent key element influencing the dynamics. 

Clearly, a strategy of motion reflecting the complex behavior of a pedestrian cannot be determined by 
a unique feature within a model. Several aspects should be included in order to define the
use of space and the attitude of the pedestrian in conflicts or competitions for space. 
Part of the object of this work is to understand how some of the different elements which may conform 
a strategic profile affect the dynamics of an evacuation process when considered separately. 
This means, to disentangle such different components, and move towards a better understanding of 
the emerging results. 

With this in mind we develop a spatially-continuous discrete-time model for pedestrian dynamics 
that enables for the analysis of the influence of the individual attitudes at different stages 
of the evacuation dynamics. These stages include the decisions made at the presence of available 
space that promotes a favorable move, and the strategy adopted if confronted with conflicts 
that may derive from these decisions. More specifically, the model considers two different stages 
at which the attitudes of the pedestrians can be introduced. First, a stage at which each pedestrian 
decides its attempt of motion according to the preferred direction, the available free space 
perceived, and reflecting its degree of rationality, patience, cooperativeness or aggressiveness. 
Second, a stage at which the conflicts among individuals that attempt to move to overlapped 
positions are solved. This second stage could in principle include Game Theory elements to reflect 
the character of the pedestrians and their fitnesses to win the competitions. However, as a first 
approach, in this work we solve the conflicts at random and 
focus on the analysis of the aspects involved in the first stage. At such a first level, the actual 
behavioral patterns of pedestrians, which may include cooperative or egotistic attitudes and rational or impulsive 
decision making, are incorporated in the model through different deterministic or noisy ways of deciding 
the attempts of motion. The selfish or cooperative attitude is mimicked at this level through an 
aggressive or coordinated 
management of the available space to move. In particular, we focus on the analysis of the effects of the {\em randomness} 
and the {\em rationality} in the use of empty space to move, with a connotation that will become 
clear later. We also study the influence of the minimal separation threshold considered by the 
pedestrians for 
attempting a step. This parameter is related to the inclination of the individuals to avoid 
collision with other pedestrians and, thus, to their degree of cooperativeness or politeness.

In agreement with the suggestions in \cite{mous}, the proposed model starts from very simple 
assumptions and a small number of parameters. Moreover, given that it considers finite-size 
pedestrians in a continuous-space, it can be thought as more realistic than most cellular automaton 
models. Furthermore, as mentioned before, it enables for the inclusion of Game Theory formalism
to solve the conflicts which emerge in the competition for space. Due to these properties, we 
believe that the model can be relevant for the analysis of many aspects of pedestrian dynamics 
beyond those studied in this work. Moreover, by considering appropriate adaptations and more complex 
geometries, it may constitute a useful tool for the design of actual spaces and buildings.

\section{The model}

We consider a spatially continuous model of $N$ pedestrians, represented by 
rigid disks of radius $r=d/2$, occupying a square room of size $L$ with a single 
door. In $(x,y)$-coordinates, the walls are located at $x=\pm L/2$, $y=0$ and 
$y=L$, while the exit is in the middle of the $y=0$ wall, running from $(-l_d/2,0)$ to $(l_d/2,0)$. 
The time is discretized in steps of duration $\delta t=1$ (in arbitrary units) 
while the space is continuous so that the position of the center of the disks 
can be any point in the room.
We analyze two versions of the same basal model, that will be referred to as 
\emph{rational} and \emph{stochastic}. In both cases, the individuals head 
towards the door in order to leave the room. More precisely, a walker located at 
position $\vec{r}=(x,y)$ heads towards the point $\vec{r}_E$ at the exit, which 
is defined as $\vec{r}_E=(0,0)$ if $|x|\ge l_d/2$ or as $\vec{r}_E=(x,0)$ 
if $|x|< l_d/2$. 

\begin{figure}
\includegraphics[width=\columnwidth]{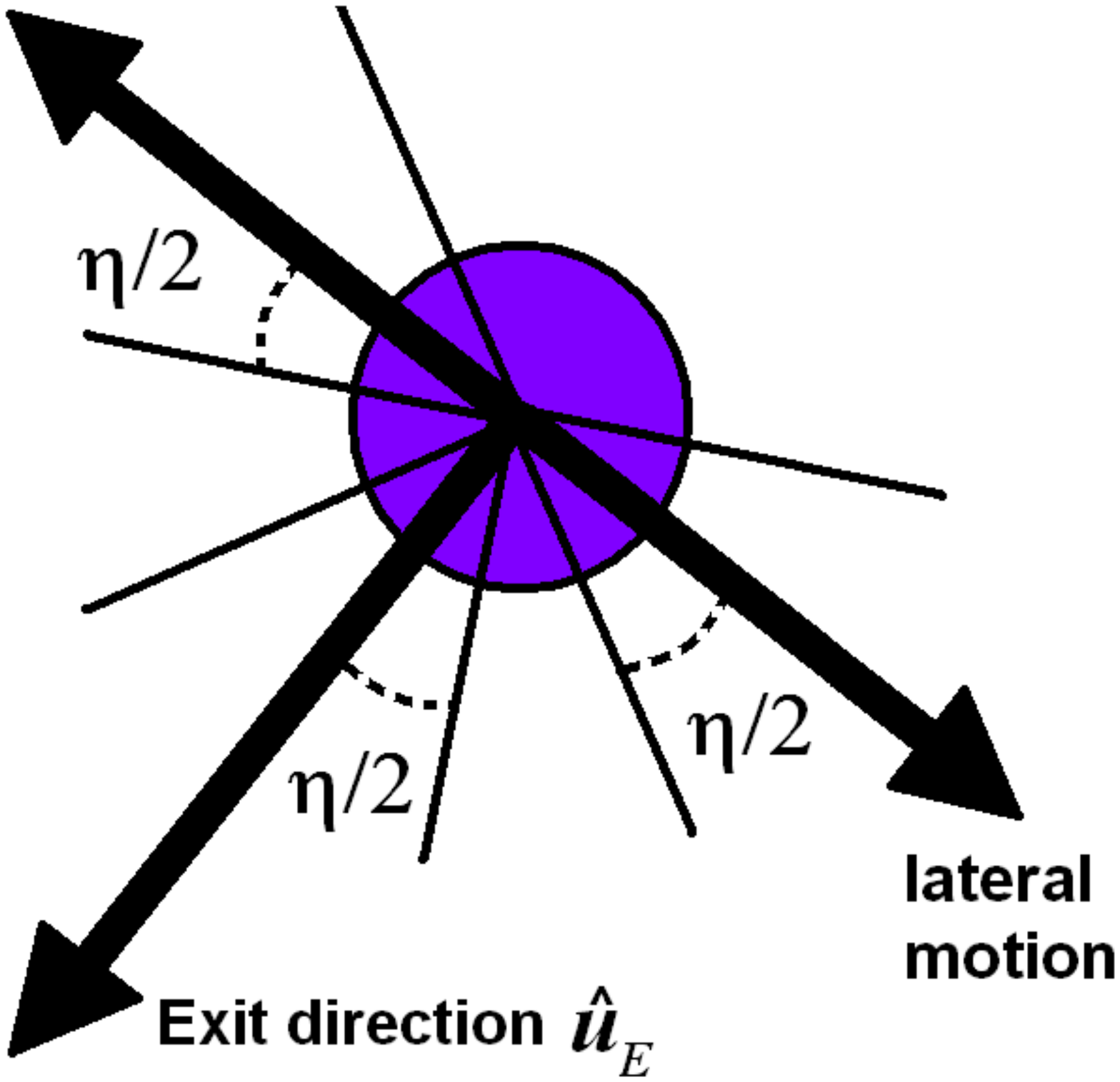} 
\caption{Scheme showing the direction of frontal and lateral movements and the uncertainty 
introduced by $\eta$}
\label{esque}
\end{figure}

The formulation for the rational model is the following.
At each time step, each walker selects a {\em desired position} to which he/she will attempt to 
move. 
For this, the walker first explores the possibility of moving in the direction of the exit, 
and if it is not possible, he/she explores the possibility of making a lateral motion. This 
is done with the following algorithm. Let $\vec{r}_i$ be the position of the walker $i$ at time 
$t$, which is represented by the center of the circle in the sketch in Fig. \ref{esque}. 
We define $\hat{u}_E=(\vec{r}_E-\vec{r}_i)/|\vec{r}_E-\vec{r}_i|$ (i.e. the direction 
towards $\vec{r}_E$). As a first attempt, the walker selects a direction of 
motion $\hat{u}$, which is randomly chosen within an angle of amplitude 
$\eta$ around $\hat{u}_E$ (i.e. $\eta/2$ to each side), as shown in Fig. \ref{esque}. Note 
that this implies that we consider an uncertainty or randomness in the way the pedestrians 
point toward the exit, which is characterized by the parameter $\eta$. 
Then, the walker analyzes what is the maximum distance he/she can move in the 
direction $\hat{u}$ without overlapping another walker 
(for this we consider all the walker's positions at time $t$). If such a 
distance is larger than a threshold $\mu\, d$  ($0<\mu<1$), the desired position for the walker 
$i$ is set as $\vec{r}_i+\lambda \hat{u}$. Here, $\lambda>\mu\, d$ is 
the size of the desired step, which is set as long as possible up to a maximum 
distance $d$ (i.e. $\mu\, d<\lambda<d$). Conversely, if the maximum distance 
allowed in the direction $\hat{u}$ is smaller than $\mu\, d$, the walker attempts 
for a lateral motion (see Figure \ref{esque}). For this, he/she randomly selects 
one of the two directions perpendicular to $\hat{u}$, here referred to as 
$\hat{u}_{\perp}$. Then, the walker checks the maximum  distance allowed for a 
motion in the direction $\hat{u}_{\perp}$. If this results larger than $\mu\, d$, 
the desired position is set as $\vec{r}_i+\lambda \hat{u}_{\perp}$. Instead, if there is a
blockage, the walker will not attempt for moving in the considered time step, 
and he/she will remain at position $\vec{r}_i$ at time $t+\delta t$.

Once all the individuals have defined the positions for their motion attempts, 
we have to check for the possibility of conflicts. Note that a conflict may 
arise when the desired positions of two or more walkers overlap. In such a case, 
the conflict is solved by selecting at random one of the walkers involved, which 
will be finally allowed to move. The rest of the individuals in the conflict 
will remain at their original positions. Note that, according to what was indicated 
in the introduction, this is the stage of the model at which Game Theory could be 
included to solve the conflicts, as done for instance in \cite{bou}.

It is worth remarking that the rational model includes two main parameters defining the 
stepping dynamics. Namely, the angle $\eta$ characterizing 
the randomness in the pointing toward the exit by the walkers, and $\mu$ 
($0<\mu<1$), which defines the threshold distance ($\mu d$) for attempting a step. 

Now we introduce the stochastic model, whose main difference with the rational 
one is that the attempt for a lateral motion is decided at random instead of being a 
consequence of the impossibility of making a forward motion. 
The algorithm is the following. At each time step, for each walker, 
the directions $\hat{u}$ and $\hat{u}_{\perp}$ are defined using $\hat{u}_E$ 
and $\eta$, just as in the deterministic case. Then, each walker is set to attempt for a 
lateral motion (direction $\hat{u}_{\perp}$) with probability $\alpha$ or for a forward motion 
(direction $\hat{u}$) with probability $(1-\alpha)$. If the available distance in 
the chosen direction is larger than $\mu d$, the walker will attempt for a step of 
length $\lambda$ in such direction. In contrast, if there is a blockage, the walker will 
not attempt for moving. After all the individuals have defined their desired positions, 
the conflicts are solved in the same way as for the rational model. 

Two things should be stressed. First in addition to the parameters $\eta$ and 
$\mu$ of the rational model, the stochastic model has a third 
parameter $\alpha$ ($0\le\alpha\le 1$), which defines the probability of attempting a
lateral movement. Second, the random choice between the directions $\hat{u}$ and $\hat{u}_{\perp}$
reflects an {\em irrational} behavior of the pedestrian. Note that the available distance is checked
only in the randomly chosen direction. 

For both models, as initial condition we consider random positions for the $N$ pedestrians 
within the room, taking care of avoiding overlapping. 

The parameters $\mu$, $\eta$, and also $\alpha$ for the stochastic case, define the 
way in which the pedestrians decide their attempts of motion. Thus, they can be interpreted as 
dependent on the characters of the pedestrians. For instance, a relatively large value of $\mu$ 
could be related to a more respectful (cooperative) attitude than
a low value of $\mu$, since a walker with larger $\mu$ is more careful in 
avoiding physical interaction with the pedestrians in the direction he/she 
moves. Similarly, larger values of $\eta$ or $\alpha$ indicate a lower propensity for moving 
straight to the exit, so that this may be also attributed to a more cooperative attitude.
The predisposition of the individuals can be
naturally introduced in the model to define the strategy of motion, that affect
both the decision making process when searching for empty space to move
(ruled by the parameters $\mu$, $\eta$ and $\alpha$), and the attitude or strategy during a
conflict resolution. As indicated in the introduction, in this work we focus on
the effects observed only during the first case. Our definition of cooperative
and defective attitudes alludes to the observed behavior during the choice of
the next move and not to deeper internal motivations that the individuals
may have. In this sense, walking straight to the exit ignoring the presence
of the others is the most intuitive and obvious interpretation of an egotistic
attitude.

\section{Results}

\subsection{Exit times and dependence on the parameters.}

\begin{figure}
\includegraphics[width=\columnwidth]{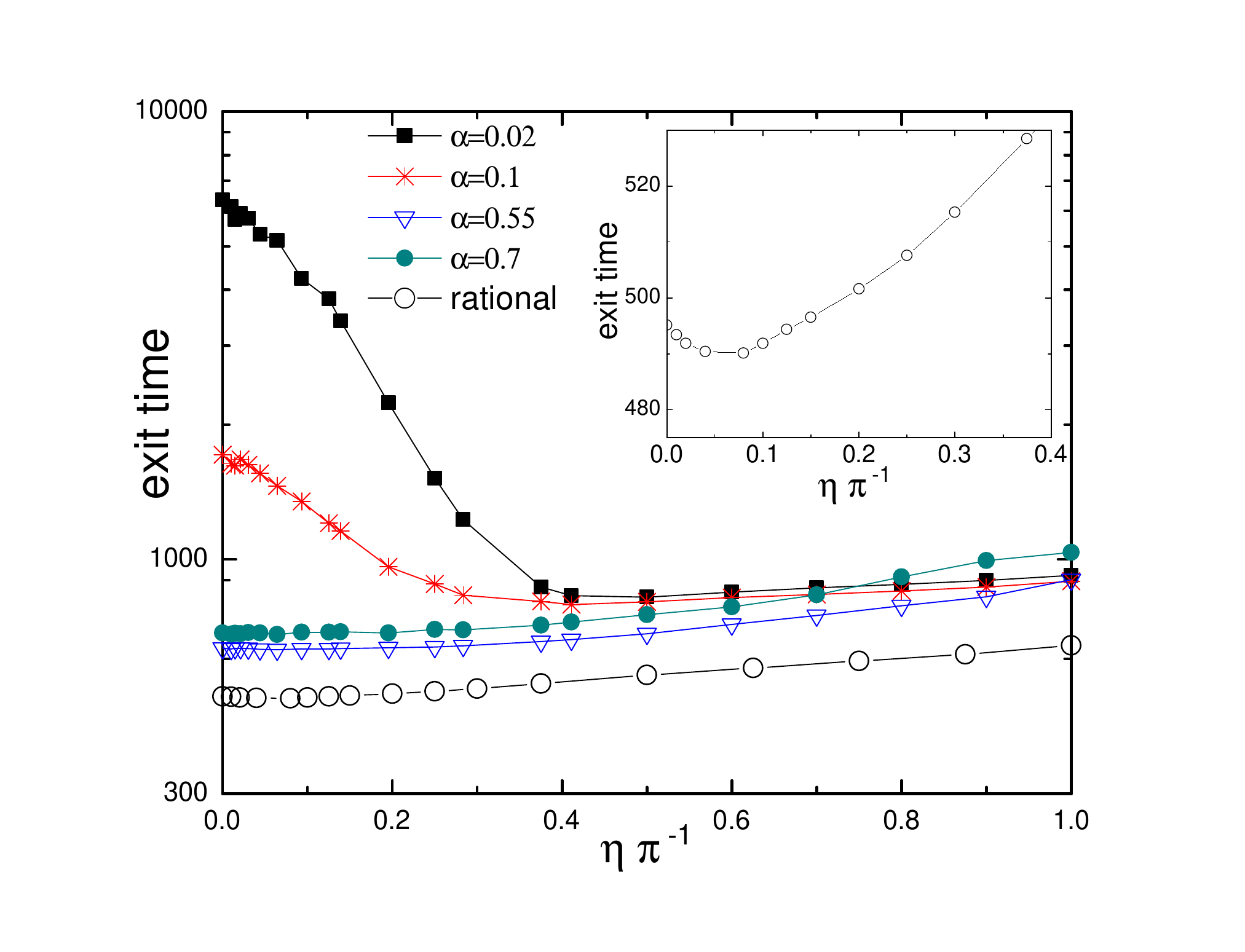} 
\caption{Mean evacuation time for the rational and stochastic models as a function of 
$\eta/\pi$. Here each curve corresponds to a different choice of $\alpha$ with 
$\mu = 0.1$, while the solid black curve with white circles corresponds to the rational 
version of the model, zoomed in the inset. The results were 
obtained by averaging over 1000 realizations, considering a room of size $L = 
100$ with a door of size $l_d = 6d$ filled to a 40\% of its capacity, containing 
$N=1000$ agents.}
\label{stoc1}
\end{figure}

We begin by analyzing the dependence of the mean evacuation time (or exit time) on $\eta$
within the two models. For this we consider a fixed value $\mu=0.1$  and 
different values of $\alpha$ in the case of the stochastic model. The effects of varying 
that threshold $\mu$ will be studied later. The results are shown in Figure \ref{stoc1}. 

First we observe that, for any value of $\alpha$, the stochastic model results in longer  
exit times than the rational one. As we show later, this result is independent
of the values of the rest of the parameters with the exception of very small values of 
$\mu$. We can affirm that, in most cases, the rational behavior  is strategically more  effective 
than the stochastic one (both for each individual and for the crowd). 

Figure \ref{stoc1} also show us that the evacuation time has a non monotonic behaviour, adopting a 
minimal value for an intermediate value of  $\eta$. This indicates that  the presence of certain 
limited amount of \textit{noise} improves the overall performance of the pedestrians, but 
the effect is reversed once $\eta$ exceeds a critical value, i.e., with an 
increase in the uncertainty when choosing the moving direction. This can be understood by 
considering that for very small $\eta$, the desired directions of movement of several pedestrians 
can point to similar targets, increasing the possibility of blockage, while large uncertainties can 
lead to unnecessary winding of the trajectories increasing the exit time. 
We have verified the robustness of this result against changes in $\mu$, as shown later, 
in the sizes of the door and the room and in the number of pedestrians.

\begin{figure}
\includegraphics[width=\columnwidth, clip=true]{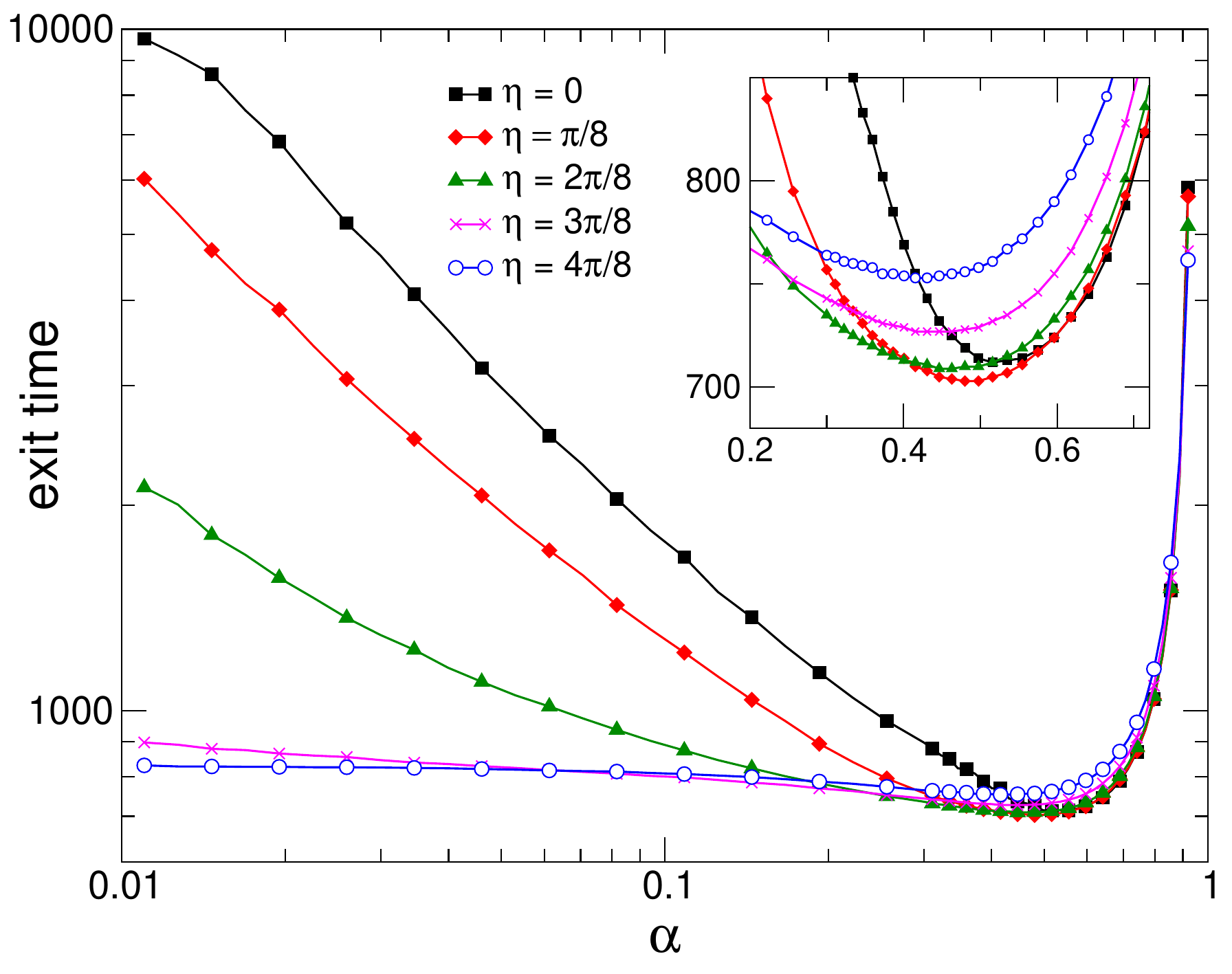}
\caption{Mean evacuation time for the stochastic model as a function of $\alpha$ with 
$\mu=0.1$. We consider different values of $\eta$. The results were 
obtained by averaging over 1000 realizations, considering a room of size $L = 100$ with a 
door of size $l_d = 6d$ filled to a 40\% of its capacity, containing $N=1000$ 
agents. The inset shows a zoom of the curve corresponding to the rational case}
\label{stoc2}
\end{figure}

Additionally, we observe that for the stochastic model there is an optimal non-negligible value of 
$\alpha$ that minimizes the exit time. To observe this more clearly we present in Figure 
\ref{stoc2} the information about exit times for different values of $\eta$ as a function of 
$\alpha$. Qualitatively, the dependence of the exit time on $\alpha$ has the 
same behavior for any value of $\eta$. There is always an optimal value that 
minimizes the exit time and that is almost constant for a given size of the 
door. For example, for  $l_d=6d$, the optimal value is $\alpha \approx 0.47$ with only 
small variations (less than 7.5\%) as a function of $\eta$, as shown in the 
inset of Figure \ref{stoc2}.

\begin{figure}
\includegraphics[width=\columnwidth, clip=true]{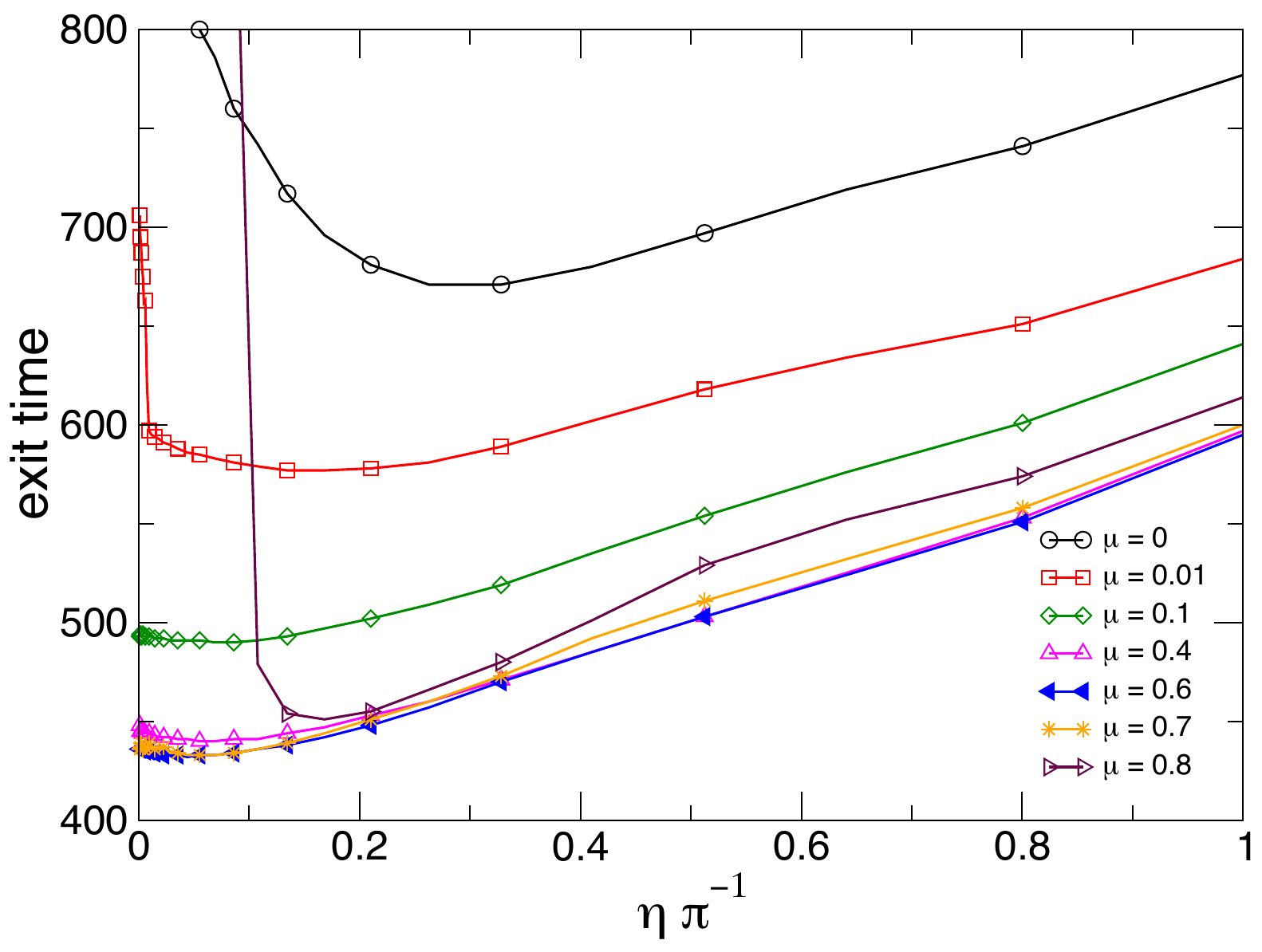}
\caption{Mean evacuation time for the rational model as a function of $\eta/\pi$ 
for different values of $\mu$. The results were obtained by averaging over 1000 
realizations, considering a room of size $L = 100$ with a door of size $l_d = 
6d$ filled to a 40\% of its capacity, containing $N=1000$ agents.}
\label{det1}
\end{figure}

For that optimal situation, the resulting exit time is of the order of the one 
found for the rational model, and even the dependence on 
$\eta$ is qualitatively the same, as shown in Figure \ref{stoc1} for stochastic and 
rational agents. However, this is not the case for lower and 
higher values of $\alpha$. On the one hand, when $\alpha$ is very low, the whole 
group of pedestrians might end up in a clogged configuration or, with some 
intermittences, stay blocked until a noisy event helps to clear the blockage. 
We have observed these two effects in several simulations. 

\begin{figure}
\includegraphics[width=\columnwidth]{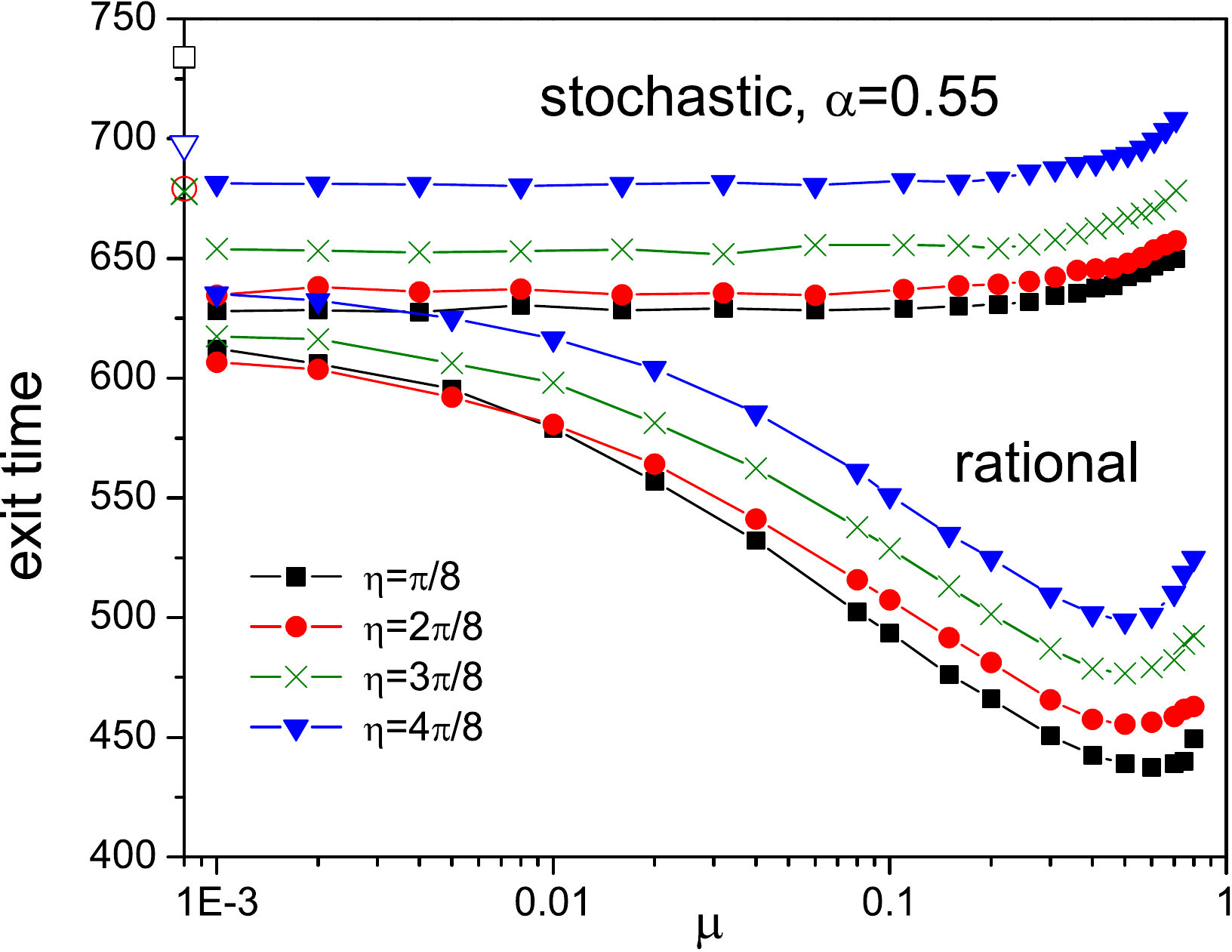}
\caption{Mean evacuation time as a function of $\mu$. Here each curve 
corresponds to a different choice of $\eta$. We show the rational and the 
stochastic cases, for the last one, the case when $\alpha$ is optimum. The 
horizontal axis is in units of $d^{-1}$. The results were obtained by averaging 
over 1000 realizations, considering a room of size $L = 100$ with a door of size 
$l_d = 6d$ filled to a 40\% of its capacity, containing $N=1000$ agents.}
\label{umbral}
\end{figure}

\begin{figure}
\includegraphics[width=\columnwidth]{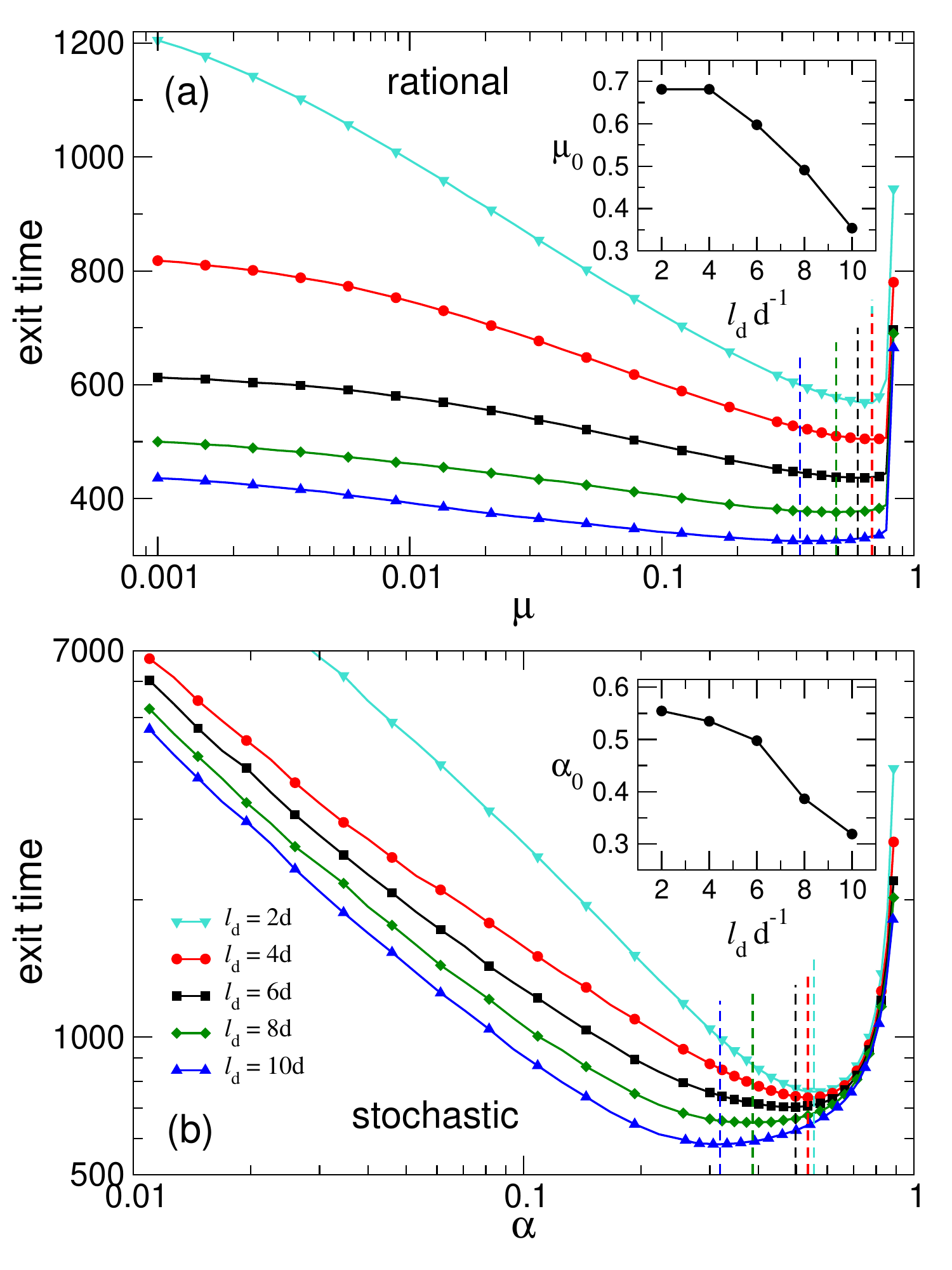}
\caption{Mean evacuation time a) for the rational case and as a function of $\mu$, b)
for the stochastic case as a function of $\alpha$. The results were obtained by averaging over 1000 
realizations, considering a room of size $L = 100$  and different  door sizes, as shown in b). 
The room contained $N=1000$ agents.}
\label{door}
\end{figure}

As mentioned before, the increase in the exit time is mainly due to a noisy 
wandering behavior (more evident for the stochastic version of our model) or to 
the occurrences of blockages due to the lack of space for the pedestrians to 
move. The first can be tuned by the amount of noise considered in the dynamics (characterized
by the parameters $\eta$ and $\alpha$). This, in turn, also \hbox{affects} the 
distribution of the available space and consequently the number of blockages. Still, there 
is something else we can do to prevent clogging. Namely, to increase the parameter 
$\mu$, which characterizes the minimal empty space that a given individual has to 
observe for attempt a motion in certain direction. 

\begin{figure*}
\includegraphics[width=\textwidth]{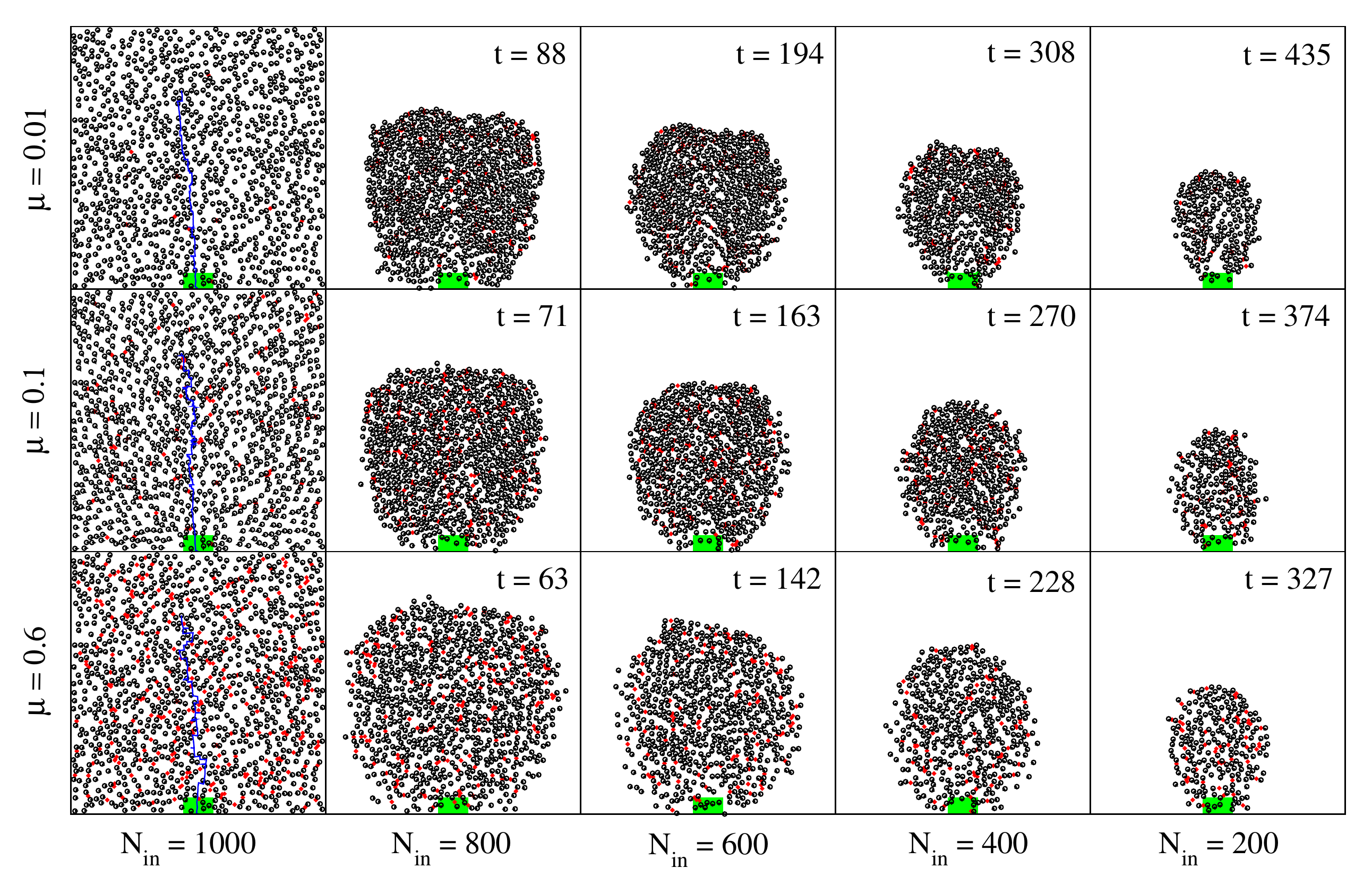}
\caption{Snapshots for different instances of the evacuation process with 
\emph{rational} agents (black circles) for different values of $\mu$ and $\eta = 
\pi/8$. The red dots represent the conflicts between two or more agents that are 
solved randomly with just one them advancing to the desired position. The green 
square represents de position and width of the door at the bottom wall of the 
room with $L = 100$ and $l_d = 6d$. The initial condition on the left column is 
the same in all cases, with the room filled to a 40\% of its capacity with 
$N=1000$ agents, while the blue curve corresponds to the trajectory of one of 
the agents until it leaves the room.}
\label{det_ss}
\end{figure*}

\begin{figure*}
\includegraphics[width=\textwidth]{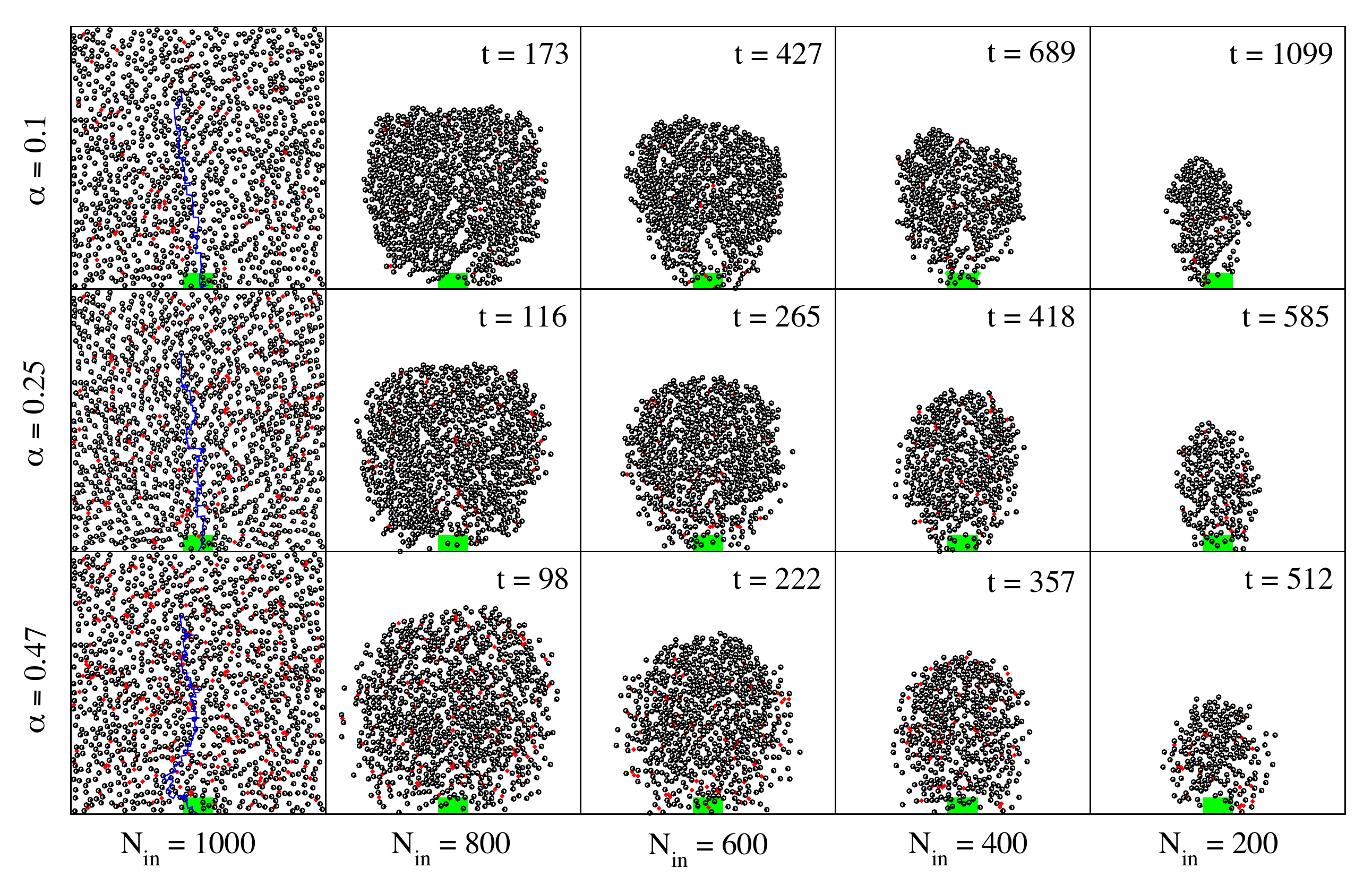}
\caption{Snapshots for different instances of the evacuation process with 
\emph{stochastic} agents (black circles) for different values of $\alpha$, $\mu 
= 0.1$, and $\eta = \pi/8$. The red dots represent the conflicts between two or 
more agents that are solved randomly with just one them advancing to the desired 
position. The green square represents de position and width of the door at the 
bottom wall of the room with $L = 100$ and $l_d = 6d$. The initial condition on 
the left column is the same in all cases, with the room filled to a 40\% of its 
capacity with $N=1000$ agents, while the blue curve corresponds to the 
trajectory of one of the agents until it leaves the room.}
\label{stoc_ss}
\end{figure*}

In Figure \ref{det1} we show the exit-time vs. $\eta$ curves for several values of $\mu$ 
within the rational model. We observe that the effect of preserving certain distance from 
the other pedestrians favors the evacuation speed unless this distance is too large and so the 
restriction has a completely different effect: the effective free space to move is reduced 
to the point that no one can move. As a consequence, the dependence of the exit time on 
$\mu$ shows that there is an optimal value ($\mu\approx 0.6$ for a door of size 
$l_d=6d$). When $\mu> 0.75$ there is a considerable proportion of realizations 
that show transient or permanent blockages. Surprisingly, this phenomenon is not 
replicated when considering the stochastic model, for which increasing the value of $\mu$ 
results always in a longer exit time, as shown in Figure \ref{umbral}. This fact 
might be indicating that the randomness in the lateral motion (measured by $\alpha$) 
is already efficient in creating empty spaces and adding the effect of $\mu$ results 
also in an effective reduction of the available space as $\mu$ increases.
Although Figure \ref{umbral} only shows the exit time for $\alpha=0.55$, we have studied what 
happens for several other values of $\alpha$ observing in all the cases the same 
behavior. 

We must mention that the optimal value for $\mu$ in the case of rational agents 
(see Figure \ref{umbral}), or for $\alpha$ in the case of stochastic ones (see 
Figure \ref{stoc2}), depends on the size of the door $l_d$ as shown in Figure 
\ref{door}. In particular, the insets in the figure shows the dependence of the 
position of the minimal, $\mu_0$ and $\alpha_0$, as a function of $l_d$ for the 
rational and stochastic cases, respectively. As apparent from the figure, these 
results suggest that, in order to optimize the flow of pedestrians out of the 
room in an evacuation situation, one needs only to control the mean distance 
among the pedestrians in relation to the size of the door itself (and the mean 
diameter of the agents): the narrower the door, the more distance is required 
among agents to keep an optimal flow through the door. 

Now let us go back to the interpretations of large and small values of the parameters $\mu, \eta$ and $\alpha$ 
as signatures of cooperative or egotistic behavior, respectively. From this point of view, the fact that 
we find non vanishing values of these parameters minimizing the exit time indicates
that neither the most egotistic nor the most cooperative behavior of individuals is optimal from the perspective
of the crowd. In the following section we analyze the connection of this results with a well known effect observed
in evacuation processes.

\subsection{Analogies with the \emph{faster-is-slower} effect}

The so called \emph{faster-is-slower} effect, originally predicted in \cite{hel2}, 
is associated to the fact that the optimal evacuation time through a narrow door is minimized at a 
given critical value at which the pedestrians try to move \cite{par07, suz13}. Hence, for desired 
velocities above the critical one, we have that the larger the  velocity, the slower 
the evacuation process. This effect has been recently confirmed experimentally for 
many different systems \cite{pas15}, suggesting that this phenomenon as a universal 
behavior for active matter passing through a narrowing.

As we now indicate, some of the simulations results presented before have clear analogies
with the {\em faster-is-lower} effect. In the present model, the speed of movement is 
not a parameter as in \cite{hel2}, it is always $d$ though the moves might be limited due to the 
lack of free space.
The haste of the pedestrian to leave the room is captured in different ways 
by the model parameters. For instance, a small value of $\mu$ indicates an impatient or hurried
behavior, since the pedestrians try to take advantage of every empty space available, without caring
about the distance to other individuals. Meanwhile, low values of $\eta$ and  $\alpha$ ( only in 
the stochastic model) indicate a propensity to walk straight to 
the exit, and a low disposition to perform lateral motions or temporal turn offs which may benefit 
the whole group.  Thus, for instance, the increase in the exit time with decreasing $\mu$ obtained
for the rational model at low values of $\mu$ resemble the {\em faster-is-slower} effect. 
The same can be said about the increasing of the exit time obtained for decreasing $\alpha$ 
observed within the random model at small $\alpha$, or the increasing with decreasing $\eta$ for 
both models at small $\eta$. It is worth saying that in the three cases  there is a region 
where the exit time decrease monotonically, until reaching the critical values at which the {\em 
faster} is {\em slower} effect starts to be observed. This monotonic behaviour has been observed 
when the the pedestrians only show a moderated haste  \cite{china}.

To understand the dynamical properties that lead to the {\em faster-is-slower-}like effect at 
small values of the parameters, we analyze in more detail the evolution of the populations and 
individual position for various simulations. In Figures \ref{det_ss} and \ref{stoc_ss} we 
present snapshots of different instances of the evacuation process for the rational and 
stochastic models, respectively, considering the same initial condition. The 
bottom row presents the optimal situation in each case for the value of $\eta$ 
and $l_d$ (and $\mu$ for the stochastic case) considered. Notice how the 
generation of available space between the agents, through the threshold 
for the rational case (associated with the parameter $\mu$) or by performing 
random lateral movements for the stochastic one (associated with the parameter 
$\alpha$), speeds up the evacuation process. 

Moreover, these results 
put in evidence an interesting fact regarding the effects of $\mu$ on the rational agents. As 
mentioned before, if a rational agent 
cannot advance towards the door a distance of at least $\mu$, it will opt for 
a lateral movement. In contrast, a stochastic agent will randomly chose a 
lateral movement depending only on the value of $\alpha$. In both cases the agent 
will stay still if it cannot advance a distance of at least $\mu$ in the chosen direction. 
Please notice that, for the stochastic agents, this threshold has been taken into account 
with $\mu = 0.1$ in the figure. Nonetheless, the rational agents require a larger value of $\mu$ 
to optimize their exit time, as their decision making process regarding the 
implementation of frontal or lateral movements is tied to this parameter, 
contrary to what happens with the stochastic agents. Therefore, this points to 
the fact that it is the implementation of a sizable amount of lateral movements 
in both cases, with the consequent generation of space among agents, the most 
important feature to optimize the exit time of the whole group.

In each of the plots in the first columns of Figs. \ref{det_ss} and \ref{stoc_ss}, we also show the 
trajectory for one of the agents (blue solid curves). It can be seen that the trajectories
for the stochastic model are more winding that those from the rational one. This leads to the 
larger exit times observed in the stochastic model.

The {\em faster-is-slower}-like effects reminds us of  {\em social dilemmas}. This 
means, situations in which there is a conflict between individual and collective interests. In a 
social dilemma, 
each individual receives a higher payoff for a socially  defecting choice than for a socially 
cooperative choice, no matter what the other individuals do, but  all individuals are all 
better off if all cooperate than if all defect. The payoff to be selfish is 
higher than the payoff to be non-selfish, but group members are worse  off if 
everyone is selfish than if everyone is not \cite{daw,lib}. In an evacuation 
scenario, the individuals would in principle always profit from moving in a faster way to the exit,
but if everybody does the same, the result is an increase in the evacuation 
time, a collective disadvantage.

\section{Final remarks and conclusions.}

Recently it was suggested that formalisms with more simple assumptions than the Social Force model could 
lead to satisfactory predictions of pedestrian behavior \cite{mous}. In that paper, the authors 
assume that the reactions of the pedestrians are mostly governed by the visually perceived 
environment and the active search for empty spaces.
The present work points in the same direction. In the models here proposed, the movement of the pedestrians 
is driven by the wish of reaching the exit and the seek for empty space to move, and limited by the 
competition for
space with other pedestrians. 

Due to their relevance, the behavioral aspects involving the character and internal motivations 
of the pedestrians has call the attention of scientist interested in modeling evacuation processes.  
The most simple but powerful approximation is to consider that there are two types of characters, prompted 
sometimes as defectors and collaborators or patient and impatient. 
Those two archetypes represent the usual conflict presented in any social 
dilemma: the compromise between the collective and the individual interest and 
the choice to favor one or the other. A given behavioral profile or strategy of motion, however, 
result from a combination of several factors that are reflected in and affect the pedestrian 
dynamics in different ways. The present work tries to disentangle these 
effects to promote a better understanding of the results.

First, our model helps us to distinguish
between the behavioral features affecting the decision making processes associated to the
selection of the next move, and those associated to the behavior at conflicts in the
competition for space. To emphasize the relevance of the distinction between these two levels of 
behavior, we can imagine
two extreme but possible attitudes of a pedestrian. First, we can think on an individual who is inpatient 
or unkind in his/her search for space, but who avoid physical contact and, thus, acts as a 
cooperative person in conflicts. This behavior could correspond to that of a disrespectful but 
coward individual.
On the other hand, we can think on an individual who is gentle in his/her search for possibilities of motion,
and tries to avoid conflicts but who becomes a strong competitor in any conflict in which he/she is forced to enter. 

Our studies focus on the analysis of the influence of different behavioral features arising on the first 
stage, i.e. on the use of empty space. At this level, the recognition of available spaces and 
the decision processes for stepping  is affected by the internal moods and 
character of the individuals. The 
consequences of these internal states on the stepping behavior are modeled through a set of independent parameters 
defining the dynamics. Namely, $\mu$, which characterizes the tendency to keep distance from others, $\eta$, which measures
the fluctuations in the definition of the preferred direction, and $\alpha$, which defines the probability of a lateral 
movement (only for the {\em random} i.e. {\em no rational} model). We have argued that low values of these parameters can
be associated to impatient or defective behaviors while relatively large values correspond to more cooperative attitudes.
Our results show that, in most cases, the exit time is minimized at intermediate values of the parameters, 
indicating that neither the most cooperative nor the most defective attitudes are optimal from the 
point of view of the crowd. We have then argued that the existence of regions of the parameters for 
which the exit time grows with the hurry or impatience of the pedestrians indicates global behaviors analogous to 
the {\em faster is slower} effect. We have shown that a thorough explanation of the slowing down of an evacuation 
process is due to an amalgam of several effects but ultimately connected to the reduction of the  
available empty space. 

As part of our studies we have shown that a distance preserving attitude favors the evacuation by inhibiting 
the clogging. Perhaps one of the most interesting implication of this results is the possibility that, 
by communicating simple basic instructions to the pedestrians, the evacuation process can be optimized. 
If people could overcome panic by having clear instructions, the behavior associated to the rational walker 
with the addition of maintaining certain distance from others would clearly help.

As a final remark, we want to emphasize the claim given in the introduction concerning the utility that 
the model here developed may have for studying general features of pedestrians dynamics, as well as for particular 
applications.

\section{Acknowledgments}
The authors gratefully acknowledge the computing time granted on the 
super\-com\-pu\-ters THUBAT-KAAL (CNS-IPICyT) and MIZTLI (DGTIC-UNAM). VD is 
grateful to V.M.\ Kenkre (UNM) for his hospitality. The authors acknowledge partial financial 
support from CONACyT and from VIEP-BUAP (Mexico) and CONICET (Argentina).


\begin{thebibliography}{99}


\bibitem{hel1} D. Helbing, I. J. Farkas, P. Molnár, and T. Vicsek, in Pedestrian 
and Evacuation 
Dynamics, edited by M. Schreckenberg and S. D. Sharma (Springer, Berlin, 2002), 
pp. 21–58.

\bibitem{hel2} D. Helbing, I. Farkas, and T. Vicsek, Nature (London) 
\textbf{407}, 487 (2000).

\bibitem{par05} D. R. Parisi and C. O. Dorso, Phys. A (Amsterdam, Neth.) 
\textbf{354}, 606 (2005).

\bibitem{lak05} T. I. Lakoba, D. J. Kaup, and N. M. Finkelstein, Simulation: 
Transactions of the Society for Modeling and Simulation International 
\textbf{81} (5), 339 (2005).

\bibitem{par07} D. R. Parisi and C. I. Dorso, Phys. A (Amsterdam, Neth.) 
\textbf{385}, 343 (2007).

\bibitem{dos15} V. Dossetti and F. J . Sevilla, Phys. Rev. Lett. \textbf{115}, 
058301 (2015).

\bibitem{kir03} A. Kirchner, K. Nishinari, and A. Schadschneider, Phys. Rev. E 
\textbf{67}, 056122 (2013).

\bibitem{wei06} Song Wei-Guo, Yu Yan-Fei, Wang Bing-Hong, and Fan Wei-Cheng, 
Physica A \textbf{371}, 658 (2006).

%
\bibitem{nic16} A. Nicolas, S. Bouzat, and M. N. Kuperman, Phys. Rev. E \textbf{94}, 022313 (2016).

\bibitem{note1} 
Note 1: Interestingly, while most of 
the literature concerning this aspect of the problem is relatively recent, closely related 
phenomenon have been studied by social psychologists more than half century ago by setting 
up experimental devices \cite{mint, kel}. 

\bibitem{bro}  R. Brown, Social Psychology (Free Press, New York, 1965).

\bibitem{col}  J. S. Coleman, Foundations of Social Theory (Belknap Press of 
Harvard University 
Press, Cambridge, MA, 1990).


\bibitem{heli1}  S Heliövaara, H Ehtamo, D Helbing, T Korhonen
Phys. Rev. E \textbf{87}, 012802 (2013).


\bibitem{heli2} H. Ehtamo, S. Heliövaara, T. Korhonen, and S. Hostikka, Adv. 
Complex Syst. 
\textbf{13}, 113 (2010).

\bibitem{hao}Q.-Y. Hao, R. Jiang, M.-B. Hu, B. Jia, and Q.-S.Wu, Phys. Rev. E 
\textbf{84}, 036107 
(2011).

\bibitem{shi} D. M. Shi and B. H. Wang, Phys. Rev. E \textbf{87}, 022802 (2013).

\bibitem{bou} S. Bouzat and M. N. Kuperman, Phys. Rev. E \textbf{89}, 032806 
(2014).

\bibitem{wang} J. Guan, K. Wang, F. Chen, Physica A \textbf{461} 655 (2016).

\bibitem{lo} S. Lo, H. C. Huang, P. Wang, and K. K. Yuen, Fire Saf. J. 
\textbf{41}, 364 (2006).

\bibitem{rapo} A. Rapoport and M. Guyer, General Systems 11, 203 (1966).

\bibitem{mous} M. Moussa{\"\i}d, D. Helbing, G.Theraulaz, PNAS \textbf{108}, 6884 (2011).

\bibitem{suz13} K. Suzuno, A. Tomoeda, and D. Ueyama, Phys. Rev. E \textbf{88}, 
052813 (2013).

\bibitem{pas15} J. M. Pastor, A. Garcimart\'{\i}n, P. A. Gago, J. P. Peralta, C. 
Mart\'{\i}n-G\'{o}mez, L. M. Ferrer, D. Maza, D. R. Parisi, L. A. Pugnaloni, and 
I. Zuriguel, Phys. Rev. E \textbf{92}, 062817 (2015).


\bibitem{china} A. Nicolas, S. Bouzat and M. N.Kuperman, in the Proceedings of the 8th
International Conference on Pedestrian and Evacuation Dynamics (PED2016). To appear. 


\bibitem{mint} A. Mintz, J. Abnorm. Soc. Psychol. 46, 150 (1951).

\bibitem{kel} H. H. Kelley, J. C. Condry, A. E. Dahlke, and A. H. Hill, J. Exp. 
Soc. Psychol. 1, 20 
(1965).


\bibitem{daw} R. M. Dawes, Annual Review of Psychology \textbf{31}, 169-193 
(1980).

\bibitem{lib}  W. B. G. Liebrand, Simulation and Games\textbf{14}, 123-138 
(1983).




\end{thebibliography}
\end{document}